# Artificial Intelligence implementation of onboard flexible payload and adaptive beamforming using commercial off-the-shelf devices

Luis M. Garcés-Socarrás[(1)], Amirhosein Nik[(1)], Flor Ortiz[(1)], Juan A. Vásquez-Peralvo[(1)], Jorge Luis González Rios[(1)], Mouhamad Chehailty[(1)], Marcele Kuhfuss[(1)], Eva Lagunas[(1)], Jan Thoemel[(1)], Sumit Kumar[(2)], Vishal Singh[(1)], Juan Carlos Merlano Duncan[(1)], Sahar Malmir[(1)], Swetha Varadajulu[(1)], Jorge Querol[(1)], Symeon Chatzinotas[(1)]

[(1)] *Interdisciplinary Centre for Security, Reliability and Trust (SnT), University of Luxembourg*
*JFK Building, 29 Avenue John F. Kennedy, Luxembourg*
*L-1855, Luxembourg*
*Email:{luis.garces, amirhossein.nik, flor.ortiz, juan.vasquez, jorge.gonzalez, mouhamad.chehaitly, marcele.kuhfuss,*
*eva.lagunas, juan.duncan, sahar.malmir, swetha.varadarajulu, jorge.querol, symeon.chatzinotas}@uni.lu,*
*{jan.thoemel, vishal.singh}@ext.uni.lu*

[(2)] *Luxembourg Institute of Science and Technology (LIST)*
*Maison de l'Innovation, 5 Avenue des Hauts-Fourneaux, Esch-sur-Alzette*
*L-4362, Luxembourg*
*Email: sumit.kumar@list.lu*

**ABSTRACT**

Very High Throughput satellites typically provide multibeam coverage, however, a common problem is that there can be a mismatch between the capacity of each beam and the traffic demand: some beams may fall short, while others exceed the requirements. This challenge can be addressed by integrating machine learning with flexible payload and adaptive beamforming techniques. These methods allow for dynamic allocation of payload resources based on real-time capacity needs. As artificial intelligence advances, its ability to automate tasks, enhance efficiency, and increase precision is proving invaluable, especially in satellite communications, where traditional optimization methods are often computationally intensive. AI-driven solutions offer faster, more effective ways to handle complex satellite communication tasks.

Artificial intelligence in space has more constraints than other fields, considering the radiation effects, the spaceship power capabilities, mass, and area. Current onboard processing uses legacy space-certified general-purpose processors, costly application-specific integrated circuits, or field-programmable gate arrays subjected to a highly stringent certification process. The increased performance demands of onboard processors to satisfy the accelerated data rates and autonomy requirements have rendered current space-graded processors obsolete.

This work is focused on transforming the satellite payload using artificial intelligence and machine learning methodologies over available commercial off-the-shelf chips for onboard processing. The objectives include validating artificial intelligence-driven scenarios, focusing on flexible payload and adaptive beamforming as machine learning models onboard. Results show that machine learning models significantly improve signal quality, spectral efficiency, and throughput compared to conventional payload.

*This work has been supported by the European Space Agency (ESA) funded under Contract No. 4000134522/21/NL/FGL named "Satellite Signal Processing Techniques using a Commercial Off-The-Shelf AI Chipset (SPAICE)". Please note that the views of the authors of this paper do not necessarily reflect the views of ESA.*

**INTRODUCTION**

Artificial Intelligence (AI) and Machine Learning (ML) rapidly advance, impacting everyday activities and highly specialized fields. Integrating automated processes, enhanced efficiency, greater precision, innovative solutions, and predictive analytics can transform conventional approaches across various domains and be increasingly utilized in space-related applications. These technologies play a vital role in autonomous navigation, spacecraft monitoring, remote sensing, satellite constellation management, and satellite communication systems, among other critical areas [1], [2].

Very High-Throughput Satellites (VHTS) are increasing their data rates in satellite communications. However, the distribution of coverage across service areas often lacks uniformity. This inconsistency leads to capacity deficits in some beams that fail to meet traffic demands while others exceed requirements, resulting in inefficient resource management [3], [4].

This challenge can be mitigated through the integration of machine learning techniques. Methods such as interference management, signal modulation, encoding strategies, communication protocols, flexible payload configurations, and adaptive beamforming enable more efficient allocation of satellite resources by dynamically adjusting payload capacities to match demand [2], [5].

Currently, most onboard processing applications in space systems rely on space-certified General-Purpose Processors (GPPs), expensive Application-Specific Integrated Circuits (ASICs), or Field-Programmable Gate Arrays (FPGAs). These space-certified components are specifically engineered to support harsh radiation environments. However, they undergo an extensive certification process that takes several years to complete. This prolonged timeline often results in spacecraft and satellites being designed with legacy components that become obsolete as technology rapidly evolves [1], [6]–[8]. The increasing performance requirements for onboard processing, driven by higher data rates and growing autonomy demands, have rendered many current space-grade CPUs inadequate for modern applications.

Emerging technologies are exploring the use of space non-qualified commercial off-the-shelf (COTS) devices for onboard satellite processing despite the challenges posed by radiation effects [9]. COTS chipsets, driven by advancements in AI/ML architectures, offer cost-effective alternatives with faster development times, particularly for mass-produced LEO constellations [1], [6], [10], [11].

The GPU for Space (GPU4S) project showcases the potential of embedded Graphic Processing Units (GPUs) for space applications, with devices like NVIDIA Xavier NX and TX2 delivering superior performance and energy efficiency for parallel processing workloads [12]–[14]. Additionally, studies have shown that embedded GPUs can effectively handle infrared detection algorithms onboard [9]. Additional research examines the existing space application domains, surveying COTS and soft-IP embedded GPUs to assess computational power and address adoption challenges [15], [16]. Research by Steenari *et al.* identified a gap between radiation-hardened processors and COTS devices in terms of performance, reliability, and mission longevity [7]. Marques *et al.* demonstrated that FPGA-based COTS platforms excel at specific ML inference tasks, particularly space weather detection [17].

This article presents the Artificial Intelligence Satellite Telecommunications Testbed (AISTT), developed under the ESA project Satellite Signal Processing Techniques using a Commercial Off-The-Shelf AI Chipset (SPAICE), aimed at testing AI/ML in satellite communication payloads using consumer-grade chipsets. It explores hardware selection and the AI/ML onboard payload implementation, detailing the models' definition, the training and quantization processes, and the accuracy results. Additionally, it discusses integrating these technologies into onboard systems and presents preliminary results demonstrating the potential of AI/ML for satellite communications.

## AI/ML ONBOARD PAYLOAD ARCHITECTURE

The SPAICE project leverages a regenerative payload on a LEO satellite to implement AI-accelerated flexible payload and beam management algorithms. Regenerative payloads improve inter-satellite links, enhance spectral efficiency, and simplify user and gateway handovers. The selected application features a software-controlled satellite payload connected to a multibeam Direct Radiating Array (DRA) antenna with hybrid beamforming, enabling dynamic adjustments to beam bandwidth, power, and width [17]–[19].

The mission scenario imposes constraints on payload resources such as signal type, bandwidth, and power, which are critical in space systems with limited power, size, and computational capacity. The SPAICE payload is tailored for a 12U CubeSat, providing uninterrupted coverage across Europe from a sun-synchronous LEO orbit at 600 km altitude. The reference design includes a high-duty cycle (>50%) and a power output exceeding 100 W, with seven coverage beams, of which two are emulated for testbed simplification [2].

The payload design is adaptable to other mission scenarios, offering scalable solutions for more complex satellite systems.

### Hardware selection

The AI/ML onboard payload includes an inference system, a software-defined radio (SDR) RF frontend, firmware, and configuration software, forming a comprehensive platform for evaluating and improving the mission payload. This setup enables efficient data analysis and seamless interface integration, supporting optimization in key areas. After assessing existing COTS options, two viable approaches for a CubeSat payload testbed have been identified: integrating the inference process directly within the SDR's RF operations or deploying separate chipsets for the SDR and inference tasks. Utilizing a unified platform that combines AI/ML inference with the RF front-end simplifies system design by reducing the need to integrate multiple boards. However, while Radio Frequency System-on-Chip (RFSoC) devices can perform AI/ML tasks, their computational performance and energy efficiency are inferior compared to dedicated AI/ML architectures. Currently, commercially available SDRs do not integrate AI/ML accelerators. To address this gap, AMD/Xilinx announced its Versal AI RF family in December 2024, which aims to incorporate AI/ML capabilities into RF systems. Silicon samples and evaluation kits for this new series are expected to be available in the fourth quarter of 2025 [20].

The second alternative involves decoupling the AI/ML inference platform from the RF front-end and linking them via a high-speed serial interface. This setup allows using an AI-capable chipset alongside an SDR, effectively separating the inference process from the acquisition, transmission, and processing of RF data.

This approach is the recommended solution for the SPAICE project due to its alignment with project requirements, resource availability, and integration goals. It offers superior versatility and performance by leveraging dedicated AI/ML processing capabilities while utilizing existing RF interfaces, resulting in a more adaptable and efficient payload system. Future implementations of satellite payloads must focus on devices with high computational power and energy efficiency for onboard applications. Studies in [10] and [11] suggest the AMD Versal AI and NVIDIA Orin families as viable options, with the Versal AI Engines achieving superior performance per watt in FP32 operations (up to 221 GFLOPs/W) compared to Orin AGX (up to 88.6 GFLOPs/W). These metrics highlight the Versal AI Edge family as promising for embedded AI/ML applications.

Among the available COTS Versal AI boards, two key options were analyzed as described in Table 1. The AMD VCK190 Evaluation Kit features a Versal AI Core chip with 400 AI Engines, delivering up to 8 TOPs at 91.9 GFLOPs/W, but with a high-power consumption of 87 W, making it unsuitable for a 12U CubeSat with a 100 W power budget. Conversely, the iWave Systems W-RainboW-G57D Development Kit hosts a Versal AI Edge SoC with 34 AI Engines, achieving up to 1.9 TFLOPs at 95 GFLOPs/W, with a power consumption of around 20 W, making it a better fit for space-constrained missions.

For the SDR solution, the HiTech Global ZRF-FMC-4A4D was selected. This FMC+ add-on board integrates a 3rd-generation AMD RFSoC ZU48DR, featuring four 14-bit RF ADCs and DACs with a power consumption of up to 45 W, aligning well with the CubeSat's size and power constraints.

**Table 1. Considered AI/ML-capable and SDR COTS boards.**

| COTS Device | AI/ML | | SDR | |
|---|---|---|---|---|
| | **VCK190** | **iW-G57D** | **ZCU111** | **ZRF-FMC-4A4D** |
| Chip | VC1902 | VE2302 | ZU28DR | ZU48DR |
| Family | ACAP AI Core | ACAP AI Edge | ZUS+ RFSoC Gen 1 | ZUS+ RFSoC Gen 3 |
| CPU | 2×Cortex-A72 | 2×Cortex-A72 | 4×Cortex-A53 | 4×Cortex-A53 |
| RTP | 2×Cortex-R5F | 2×Cortex-R5F | 2×Cortex-R5 | 2×Cortex-R5 |
| AIE | 400 | 34 (ML) | - | - |
| DSP | 1968 (DSPE) | 324 (DSPE) | 4272 | 4272 |
| CLBs | 1968k | 329k | 930k | 930k |
| ADCs | - | - | 8×12 bits 4.096 GSPS | 4×14 bits 5 GSPS |
| DACs | - | - | 8×14 bits 6.554 GSPS | 4×14 bits 10 GSPS |
| Chip Power | ≈87 W | ≈20 W | 20 – 30 W | 30 – 40 W |
| Comp. Cap. INT8 | 13.6 – 133T | 3.2 – 32T | 5.44G | 5.44G |
| OPS/W | 36.8 – 91.9G | 35 – 95G | 181 – 272M | 136 – 181M |
| Comp. Cap. FP32 | 3.2 – 8T | 0.7 – 1.9T | 1.07G | 1.07G |
| FLOPS/W | 36.8 – 91.9G | 35 – 95G | 35.6 – 53.5M | 26.7 – 35.6M |
| Board Power | 180 W | 60 W | 180 W | 45 W |
| Size | 24 × 19 cm$^2$ | 12 × 12 cm$^2$ | 30 × 20 cm$^2$ | 7.8 × 6.9 cm$^2$ |

**Payload firmware**

Figure 1 depicts the proposed onboard satellite payload implementation firmware, highlighting in red the machine learning models implemented using the AI Engines in the Versal AI SoC. From the onboard computer (OBC), the payload receives the traffic demand for the coverage zone ($R$), the beam pointing angles in Azimuth ($Az$) and elevation ($El$), and the beamforming coefficients phasors ($e^{j\theta W}$).

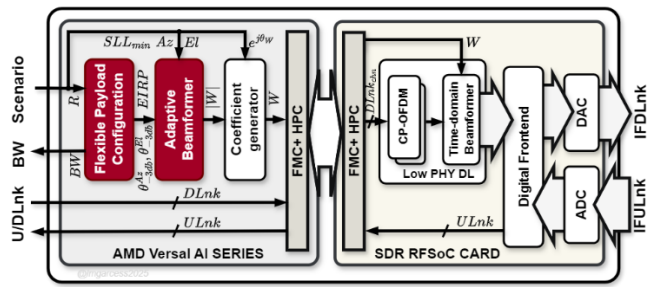

**Figure 1. Onboard Payload Firmware Diagram.**

The *Flexible Payload Configuration* module is a trained machine learning model that receives the resource configuration for the inference process and outputs the bandwidth configuration ($BW$) for the downlink signal ($DLnk$), as well as the effective isotropic radiated power ($EIRP$) and beam width in Azimut ($\theta^{Az}_{-3db}$) and elevation ($\theta^{El}_{-3db}$) per beam to the *Adaptive Beamformer*. This is also a trained machine learning model that requires the beam pointing angles (in Azimuth and elevation) and the minimum side lobe level ($SLL_{min}$), in addition to the outputs from the first machine learning model.

*Models' Definition and Training*

The *Flexible Payload Configuration*, defined in TensorFlow2 (TF2), is Kera's sequential convolutional neural network (CNN) float model with three two-dimensional convolution layers (*conv2D*) and two dense layers (*dense*) and internal operators between layers, as shown in Figure 2a. The model's input is a 401×501 matrix representing the traffic demand ($R$) depending on the satellite position. The model classifies the output into 50 classes with the proper EIRP and the bandwidth configuration for the corresponding input. On the other hand, the *Adaptive Beamforming* is also a Kera's sequential CNN float model with three dense layers (Figure 2b). The model's input is a six-element vector with the current *EIRP* per beam and satellite position on Azimut (*Az*) and Elevation (*El*). At the same time, the output is a predefined cluster for the beam coefficients configuration storage in 15 classes.

```
Model: "sequential"
_________________________________________________________________
Layer          (type)          Output Shape           Param #
=================================================================
conv2d         (Conv2D)        (None, 399, 499, 32)   320
max_pooling2d  (MaxPooling2D)  (None, 199, 249, 32)   0
conv2d_1       (Conv2D)        (None, 197, 247, 64)   18496
max_pooling2d_1 (MaxPooling2D) (None, 98, 123, 64)    0
conv2d_2       (Conv2D)        (None, 96, 121, 128)   73856
max_pooling2d_2 (MaxPooling2D) (None, 48, 60, 128)    0
flatten        (Flatten)       (None, 368640)         0
dense          (Dense)         (None, 128)            47186048
dropout        (Dropout)       (None, 128)            0
dense_1        (Dense)         (None, 50)             6450
=================================================================
Total params: 47,285,170
Trainable params: 47,285,170
Non-trainable params: 0
_________________________________________________________________
Input Shape: (None, 401, 501, 1)
```
a)

```
Model: "sequential"
_________________________________________
Layer     (type)    Output Shape   Param #
=========================================
dense     (Dense)   (None, 64)     448
dense_1   (Dense)   (None, 64)     4160
dense_2   (Dense)   (None, 15)     975
=========================================
Total params: 5,583
Trainable params: 5,583
Non-trainable params: 0
_________________________________________
Input Shape: (None, 6)
```
b)

**Figure 2. Models' definition. a) Flexible Payload Configuration. b) Adaptive Beamforming.**

The float model's training is performed using TF2 under the AMD Vitis AI tool, allowing the implementation of CNN quantized models on the AMD Deep Learning Processor Unit (DPU), a programmable engine for accelerating convolutional neural networks. The IP consists of a register configuration module, data controller module, and convolution computing module optimized by a specialized instruction set, which can be integrated into the programmable logic (PL), connected to the processing system (PS) of Zynq families, and to exploit the AI-Engines (AIE) of the Versal ACAP families [21]. Figure 3 shows the float models' losses and accuracy resulting from the training process. The *Flexible Payload Configuration* float model achieves zero loss with an accuracy of one after 50 epochs of training (Figure 3a). On the other hand, the *Adaptive Beamforming* model reports 0.0028 losses and an accuracy of 0.9989 on 18 epochs (Figure 2b).

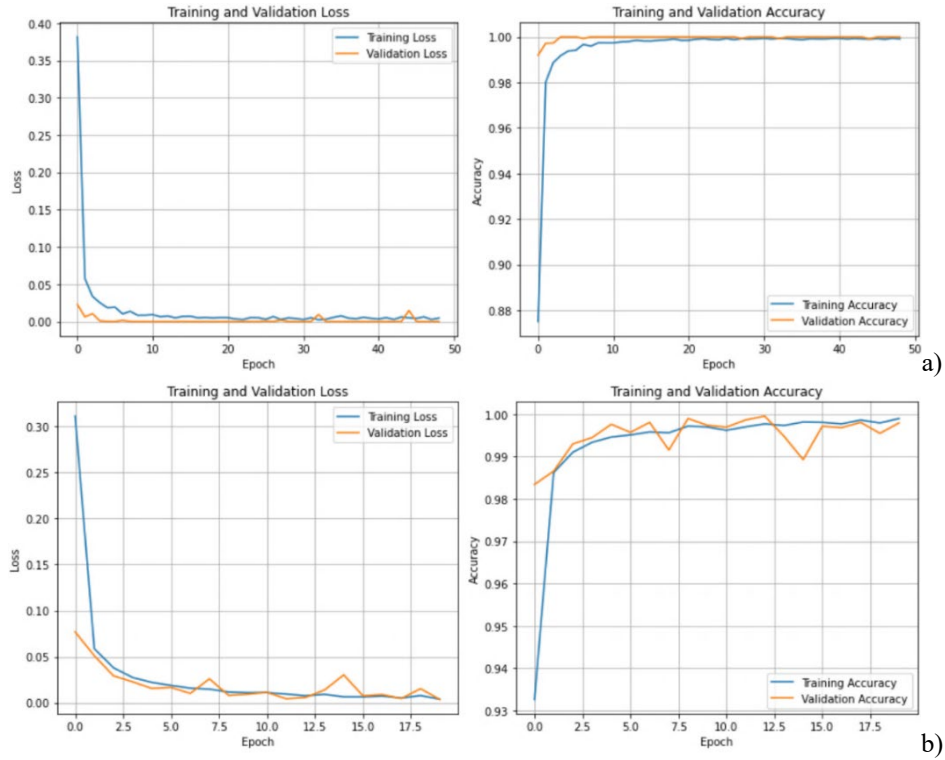

**Figure 3. Float models' loss and accuracy after training. a) Flexible Payload Configuration. b) Adaptive Beamforming.**

After the float model training, the Vitis AI allows the model's inspection to know the split of the layers and functions defined in the float model for the specific development board. The analysis of both float models reports that the two-dimensional maximum function (MaxPooling2D) on the *Flexible Payload Configuration* model and the activation function of the last dense layer (Softmax) on both models are not supported by the DPU. In these cases, those functions will be executed using the processor available on the processing system (PS) of the Versal AI Egde.

*Models' quantization*
The models must be quantized for the hardware implementation to reduce the numerical precision (from FP32 to INT8). Although this impacts the model's accuracy, it optimizes the models for size, speed, and energy efficiency, making it ideal for deployment on resource-constrained devices. It also minimizes memory usage, accelerates inference, decreases power consumption, and aligns with hardware accelerators optimized for low-precision formats. The process results of the float models' quantization are shown in Figure 4. The *Flexible Payload Configuration* quantized model maintains the accuracy achieved by the float model, as shown in Figure 4a. The losses and accuracy are slightly impacted for the *Adaptive Beamforming* quantized model, achieving 0.0348 and 0.9966, respectively (Figure 4b). Those results vary any time the process is executed, as the weight values assigned by the tool are different in each case.

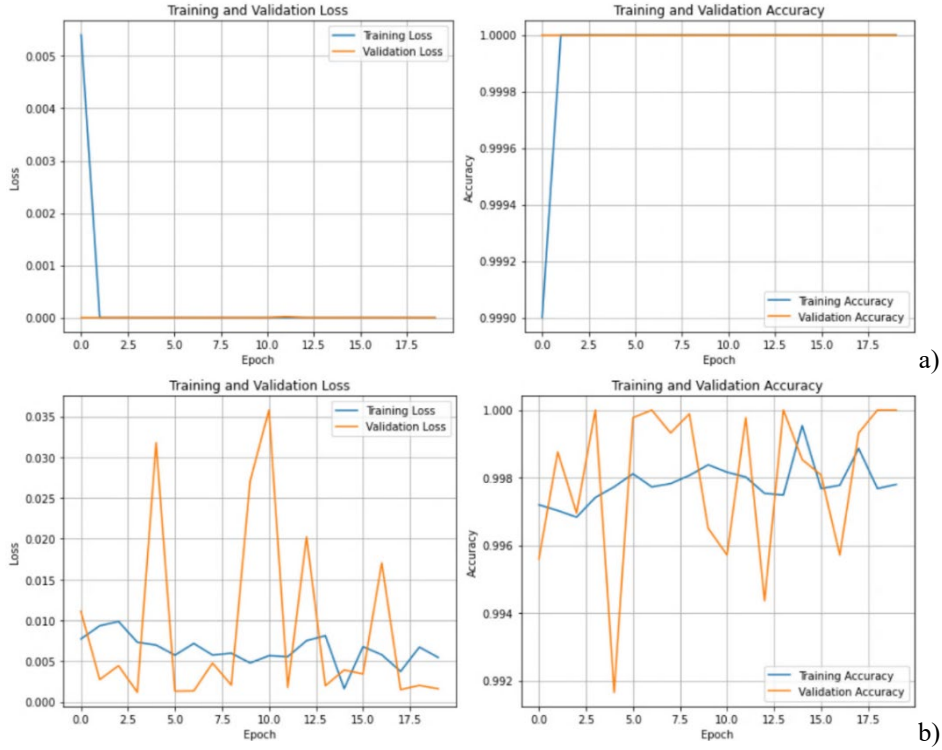

**Figure 4. Models' loss and accuracy after quantization. a) Flexible Payload Configuration. b) Adaptive Beamforming.**

*Models' execution*
The models' execution for the onboard payload is controlled by software running in standalone mode that interacts with the different elements in the design and the hardware platform design required for the DPU implementation, as shown in Figure 5. In the first step, the OBC sends the traffic demand ($R$) corresponding to the coverage area to the Versal AI Edge, which is converted to a binary file (*Input to .bin*). Later, the application invokes the execution of the *Flexible Payload Configuration* model with the binary file as a parameter. This operation runs on the DPU using the AIE for the functions supported by the IP, while the rest are executed on the PS. The resulting operation will generate another binary file that has to be imported into the application (*.bin to Output*) to continue with the operation flow. Finally, the class received from the model ($FP_{class}$) is translated to the seven beams configuration (*Beam*) and then to the EIRPs, the beam width in Azimut ($\theta^{Az}_{-3db}$) and elevation ($\theta^{El}_{-3db}$), and the required bandwidth (*BW*) by a two look-up tables (*FPayl. Class LUT* and *Beam Conf. LUT*) running on the PS.

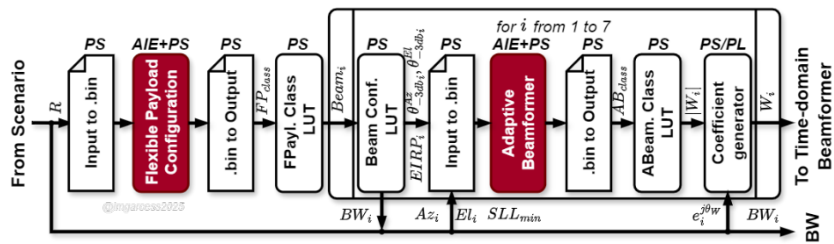

**Figure 5. Onboard Payload Machine Learning Algorithms.**

In the second step, once the EIRPs and beam widths are obtained, the configuration for one beam is converted to a binary file (*Input to .bin*) within the satellite position data and the minimum side lobe level, and the *Adaptive Beamforming* model is executed. The resulting binary output is translated to the output class ($AB_{class}$) and decoded into the beamforming coefficient modules ($|W_i|$) by a look-up table (*Abeam. Class LUT*). Finally, the beamforming modules are combined with the beamforming phasors ($e^{j\theta W}$) from the OBC on the *Coefficient generator* to obtain the beamforming coefficients for the actual beam. This second step is repeated for the total of beams, returning the beamforming coefficients ($W$) for the low-physical layer and the $BW$ for the base station.

In the low-physical layer, shown in Figure 1, the downlink modulated signals (*DLnk*) coming from the downlink source are transferred to the cyclic prefix orthogonal frequency division multiplexing (*CP-OFDM*). This function is implemented using the fast Fourier transformation (*FFT*) IP Core released by Xilinx for the RFSoC family. Once the signal is modified, the time-domain beamforming is performed by applying the beamforming coefficients exploiting the embedded multipliers in the RFSoC (*DSP48E2*). The digital front-end serves as the interface between the signals in the digital and the analog radio frequency domain. The frequency division multiplexing (FDM) and digital up-conversion (DUC) operations are applied to the output of the time-domain beamforming before it is converted to analog RF signals. At the same time, the analog RF channel signals for the uplink (*IFULnk*) are digitalized and sent back to the signal source without processing. The low-physical layer in the uplink is omitted for simplicity.

**ARTIFICIAL INTELLIGENCE SATELLITE COMMUNICATIONS INTEGRATION**

The Artificial Intelligence Satellite Telecommunication Testbed (AISTT) architecture combines AI/ML onboard payload, scenario generation, base station emulator, channel emulation, and user equipment for evaluating and optimizing the mission payload. Figure 6 depicts the AI/ML onboard payload integration with the other parts of the testbed.

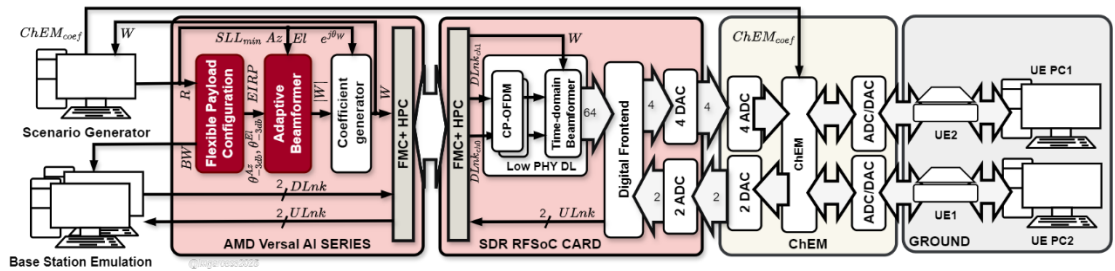

**Figure 6. AISTT Functional Diagram.**

In this case, some modifications are introduced into the onboard payload algorithm to fit the hardware constraints and emulation steps. The beamforming coefficients on the machine learning algorithm are feedback to the OBC for obtaining the channel emulator (ChEM) matrix coefficients ($ChEM_{coef}$). This is required to decode the downlink signals after mixing them in the time-domain beamformer. This operation is not computed onboard as, in a real scenario, channel emulation is not required. Due to the hardware constraints, only two downlink signals are emulated on the physical layer; four ADCs are used on the IF downlink signal and two DACs for the IF uplink.

The scenario generator is a MATLAB script within the payload control center. It produces the inputs for the AI/ML algorithm, including traffic demand ($R$), beam pointing angles ($Az$ and $El$), the minimum side lobe level ($SLL_{min}$) and the beamforming coefficient phasors ($e^{j\theta W}$). The base station generators are two Next Generation NodeB distributed units (gNB-DU) running OpenAir-Interface (OAI) that generate the two downlink signals and receive the uplinks. Two Ethernet interfaces on the Versal AI board accomplish the interconnection of the AI/ML payload and the gNB-DU computers. For implementation purposes, the partially regenerative functions running on the satellite payload are on the downlink low-PHY layer. The remaining gNB-DU functionalities, including the MAC scheduler, remain on the OAI side. Therefore, the MAC scheduler receives the instructions to adapt the beam beamwidth from the *Flexible Payload Configuration* algorithm on the Versal AI.

On the other hand, the channel emulator receives the RF signal from the AI/ML payload, applying the channel effects, and sends it to the user equipment side. It receives the signals out of the 64 antennas from the satellite DRA downlink signal multiplexed via four RF connectors that are demultiplexed at the ChEM.

In the final stage, the four RF channels of the ChEM are captured by two user equipment (UE). Such UEs are placed in different satellite beams, and they aggregate all the traffic demand corresponding to all the users served by each beam. This approach has been used in the project's simulation and training phases, and it will be mimicked in the hardware testing phase. At the same time, the OAI UE generates different user information that will be retransmitted to the gNB-DU via the ChEM and AI/ML payload (*uplink*) to modify downlink requirements.

**PRELIMINARY RESULTS**

Figure 7 illustrates the designated coverage zone, focusing on a substantial part of Europe through the orbital properties of a CubeSat equipped with seven-beam coverage. The alignment of these beams depends on the satellite's orbital location. Additionally, the figure underscores the seven beams provided for the current time step and the particular orbital transit of the CubeSat relevant to our analysis. Table 2 shows the performance requirements and results of deploying ML models onboard the payload for the reference LEO scenario. Compared to SPAICE's performance benchmarks, the signal-to-interference-plus-noise ratio (SINR) consistently exceeds 6 dB, the average spectral efficiency surpasses 1 b/s/Hz, and the throughput consistently exceeds 18 Mbps. The demand adjustment remains below 0.4, quantified by the normalized mean square error (NMSE). The maximum energy/power usage, implementation complexity, and rapid response to modifications are discussed in relation to both ML- and non-ML-based solutions, showing favorable outcomes. The focus is on the percentage increase in speed of the ML approach compared to the non-ML approach.

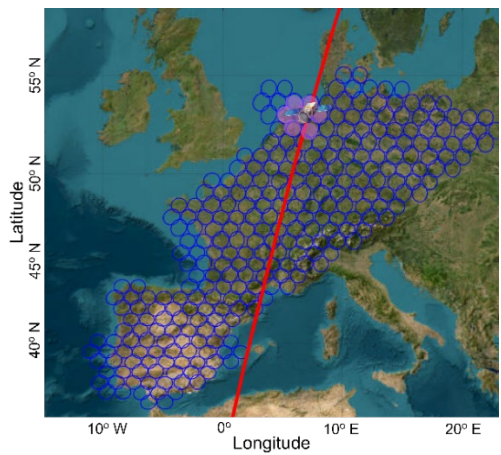

**Figure 7. Seven beams coverage area of the CubeSat trajectory over Europe** [2]**.**

**Table 2. SPAICE performance requirements and implementations for LEO Scenario**

| Description | Requirements | Flex. Payload | Adap. Beamf. |
|---|---|---|---|
| Signal-to-interference-plus-noise ratio (SINR) [dB] | > 6 | 7.553 | 7.553 – 11.35 |
| Average spectral efficiency (SE) [b/s/Hz] | > 1 | 1.3361 | 1.3361 |
| Throughput [Mbps] | > 15 | 18.66 – 110.50 | 18.66 – 110.50 |
| Demand Matching (average) | < 0.4 | 0.09 | 0.21 |
| Maximum Power. | < 40% | 6.3% | 8.6% |
| Implementation Complexity [sec] | < 60 | 3.1 | 0.682 |
| Response time | > 90% | > 97.01% | > 98.93% |

**CONCLUSIONS**

The AISTT is a useful tool for evaluating various onboard payload scenarios and the effectiveness of AI/ML techniques in satellite communication systems. Using COTS AI chipsets, the testbed offers a flexible and cost-effective solution to improve payload management and performance compared to space-qualified devices.

The solution for implementing the onboard satellite payload with an AI/ML-capable chipset and an SDR is considered the best trade-off regarding computer power per watt. The AI/ML module, powered by a Versal AI Edge chipset, performs properly and consumes less power. On the other hand, the RFSoC FMC add-on card provides RF capabilities for the payload, which consists of transmitting/receiving and modifying the RF/IF signals.

Both models achieve more than 0.99 accuracy and up to 0.035 losses after quantization. However, not all model functions will be executed on the DPU and will be carried out by the processing system in Versal.

The results show significant improvements in signal quality, spectral efficiency, and throughput, underscoring the potential for the integration of AI/ML in space applications. The AISTT's ability to simulate different mission scenarios and hardware configurations provides valuable insights, ensuring that future satellite missions can be optimized for performance and efficiency.

Beyond the current functionalities, the AISTT is designed to be scalable and adaptable to various mission scenarios, including future 6G-NTN missions. Its modular architecture supports seamless integration of next-generation chipsets and expanded payload functionalities. Future developments aim to incorporate higher computational capabilities, enhanced energy efficiency, and broader interoperability, enabling diverse applications such as full stack optimization, autonomous network management, advanced beamforming strategies, and multi-functional satellite payloads.

These capabilities offer immense opportunities for industry stakeholders to collaborate in developing tailored solutions for emerging market needs. The testbed's versatility extends to 5G/6G TN-NTN integration, space-based edge intelligence, and joint communications and sensing, providing a robust platform for experimentation and deployment. SnT is open to partner with industry leaders to leverage the AISTT and to co-develop innovative satellite technologies, accelerate product readiness, and contribute to shaping the future of satellite communication.

We invite interested organizations to collaborate and explore the AISTT's potential to drive transformative advancements in satellite communication. Together, we can redefine the capabilities of onboard processing and establish new benchmarks in performance, efficiency, and adaptability for next-generation satellite systems.